\documentclass[jgrga]{agu2001}

\usepackage{lineno}

\usepackage{graphicx}

\setkeys{Gin}{draft=false}

\authorrunninghead{USOSKIN ET AL.}

\titlerunninghead{Modulation potential for NM era}

\authoraddr{Corresponding author: I. Usoskin}

\begin{document}


\title{Heliospheric modulation of cosmic rays during the neutron monitor era: Calibration using PAMELA data for 2006--2010}




\authors{Ilya G. Usoskin,\altaffilmark{1, 2}
Agnieszka Gil,\altaffilmark{3}
Gennady A. Kovaltsov,\altaffilmark{4}
Alexander L. Mishev,\altaffilmark{1}
Vladimir V. Mikhailov,\altaffilmark{5}
}

\altaffiltext{1}{Space Climate Research Unit, University of Oulu, Finland.}

\altaffiltext{2}{Sodankyl\"{a} Geophysical Observatory, University of Oulu, Finland.}

\altaffiltext{3}{Institute of Mathematics and Physics, Siedlce University, Poland.}

\altaffiltext{4}{Ioffe Physical-Technical Institute, St. Petersburg, Russia.}

\altaffiltext{5}{National research nuclear university "MEPhI", Kashirskoye shosse 31, 115409 Moscow, Russia.}


\begin{abstract}
A new reconstruction of the heliospheric modulation potential for galactic cosmic rays is presented for the neutron monitor era,
 since 1951.
The new reconstruction is based on an updated methodology in comparison to previous reconstructions:
(1) the use of the new-generation neutron monitor yield function;
(2) the use of the new model of the local interstellar spectrum, employing in particular direct data from the distant missions; and
(3) the calibration of the neutron monitor responses to direct measurements of the cosmic ray spectrum performed by the PAMELA
 space-borne spectrometer over 47 time intervals during 2006--2010.
The reconstruction is based on data from six standard NM64-type neutron monitors (Apatity, Inuvik, Kergulen, Moscow, Newark and Oulu)
 since 1965, and two IGY-type ground-based detectors (Climax and Mt.Washington) for 1951--1964.
The new reconstruction, along with the estimated uncertainties is tabulated in the paper.
The presented series forms a benchmark record of the cosmic ray variability (in the energy range between 1--30 GeV)
 for the last 60 years, and can be used in long-term studies in the fields of solar, heliospheric and solar-terrestrial physics.

\end{abstract}



\begin{article}

\section{Introduction}

Galactic cosmic rays (GCR) is a population of energetic, mostly nucleonic, with a small fraction
 of electrons and positrons, particles permanently bombarding Earth and
 forming the radiation environment in the near-Earth space and in the atmosphere \citep{vainio09}.
The flux of GCR is modulated, in the low-energy (below 100 GeV) part of the spectrum, by solar magnetic activity
 over the solar cycle \citep{potgieterLR}.
The variability of the GCR flux is constantly monitored by the network of ground-based neutron monitors (NMs)
 since the 1950s.
Because of the thickness of the Earth's atmosphere and the shielding effect of the geomagnetic field,
 ground-based measurements have to be translated into the actual flux units beyond the atmosphere and magnetosphere
 by applying a complicated transport model.
On the other hand, GCR energy spectra are occasionally measured in the energy range exceeding 1 GeV by balloon-borne
 or space-borne detectors providing a direct way to calibrate the ground-based detectors and to link NM data to the
 real GCR spectra.
The most important in this respect are the long-running experiments PAMELA (Payload for Antimatter Matter
 Exploration and Light-nuclei Astrophysics) \citep{adriani13} and AMS-02 (Alpha Magnetic Spectrometer) \citep{aguilar15},
 operating for the last decade.
Before that, only balloon-borne detectors (and a short test flight of AMS-01 in 1998 \citep{alcaraz00}) were operating
 in this energy range.

For many practical purposes it is ueful to describe the GCR energy spectrum near Earth by the force-field approximation
 \citep[e.g.][]{gleeson68,caballero04} with its single formal parameter -- the modulation potential $\phi$ \citep[see formalism in][]{usoskin_Phi_05}.
We note that the force-field approximation is not validated as a physical model of GCR modulation, and the
 modulation potential has no clear physical meaning
 (often used interpretation of the mean adiabatic energy loss is not exactly correct \citep[see, e.g.][]{caballero04}).
On the other hand, it provides a handy empirical description of the actual shape of the GCR energy spectrum near Earth which,
 while not making a claim to explain the modulation process, offers a simple single-value parametrization of the
 GCR spectrum for many practical purposes, such as atmospheric ionization and climate modeling, radiation environment,
 cosmogenic radionuclide studies, assessments of radiation hazard risks, etc.
A model allowing one to estimate the variability of the modulation potential in time was proposed by \citet{usoskin_Phi_05}
 based on the data from the world NM network.
That work led to a systematic reconstruction of monthly $\phi$ values since the 1950s.
Calibration to the direct GCR measurements was done using the space-borne  AMS-01 data for moderate solar activity and
 MASS89 balloon-borne data \citep{webber91} for high solar activity.
This work was extended by \citet{usoskin_bazi_11} by including a more realistic GCR composition (heavier species were considered).

Here we revisit the reconstruction of the modulation potential along three main directions:
\begin{enumerate}
\item
The earlier models were based upon previous generations of the NM yield functions \citep{debrunner82,clem00,matthia09} that were unable
 to reproduce the exact count rate of individual NMs and the shape of the latitudinal survey \citep{caballero12}.
 By contrast, here we use the new-generation NM yield function \citep[][see also erraturm therein]{mishev13}, which agrees,
  for the first time, with the actual measurements of the NM count rates and observational surveys \citep{gil15}.

\item
While the earlier models were based upon an estimate of the local interstellar spectrum (LIS) by \citet{burger00}
 for earlier models such as \citep{garcia75},
 here we use a recent estimate of the LIS by \citet{vos15} who revised the LIS by using precise measurements from AMS-02 and PAMELA
 space-borne detectors and considering also Voyager data beyond the heliospheric termination
 shock, not available until recently.

\item
Earlier models were based upon a calibration method using only two directly measured GCR spectra: MASS89 and AMS-01.
Here we use a newly available GCR spectra precisely measured by the PAMELA instrument \citep{adriani13} during 47 time intervals
 during 2006 -- 2010.
\end{enumerate}

We note that with these modifications (especially the new LIS and calibration), the values of $\phi$ calculated here
 are not directly comparable with the earlier reconstructions.

In Section~\ref{Sec:model} we describe the formalism of the model and the used LIS.
The PAMELA data used for calibration are introduced in Section~\ref{Sec:PAMELA}.
Selection and calibration of the NMs are explained in Section~\ref{Sec:calib}.
The reconstruction of the modulation potential is described in Section~\ref{Sec:recon} and discussed
 in Section~\ref{Sec:result}.
Our conclusions are presented in Section~\ref{Sec:conc}.


\section{Formalism}
\label{Sec:model}
Here we use the established formalism of representing the counting rate of a NM at
 any location and time $t$, as an integral of the product of the cosmic ray energy spectrum and the specific yield function of the NM:
\begin{linenomath*}
\begin{equation}\label{Eq:N}
  N(h,t)=\kappa\sum_{i}\int_{T_{c_{i}}}^{\infty}Y_{i}(T,h)\cdot J_{i}(T,t)\, dT,
\end{equation}
\end{linenomath*}
where $N$ is the count rate of a NM reduced to the standard barometric pressure, $J_i(T,t)$ is
 the energy spectrum of the $i-$th specie of GCR nuclei outside the Earth's magnetosphere and atmosphere, $Y_i(T,h)$ is the
 specific yield function of a NM, $T$ is kinetic energy of the primary cosmic rays particle, $h$ is height (atmospheric
 depth at the NM location), and $\kappa$ accounts for the ``non-ideality'' of a NM (see Section~\ref{Sec:calib}).
The yield function, corresponding to the standard sea-level 6NM64, was taken according to a recent simulation
 \citep[][see also erratum therein]{mishev13}.
Integration is performed above the kinetic energy $T_c$ corresponding to the geomagnetic cutoff rigidity $P_c$ in the location of
 the NM.
The yield function includes both development of the atmospheric cascade with different types of secondary particles
  and the response of a detector to the secondary particles \citep{clem00,mishev13,aiemsa15}.

\begin{figure}[th]
\centering
\includegraphics[width=\columnwidth]{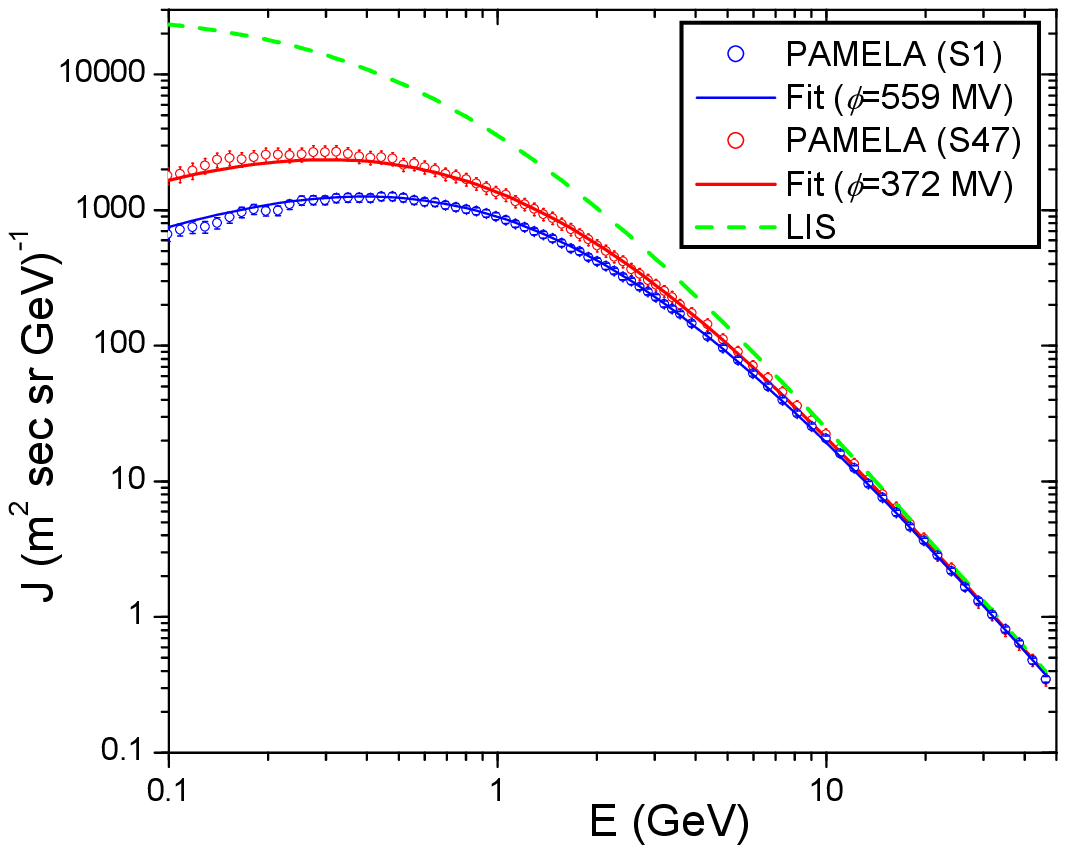}
\caption{Examples of the fits (curves) of the GCR proton spectra measured by PAMELA (dots with errors bars) for
  period S1 (07.07.2006--26.07.2006, blue) and S47 (02.01.2010--23.01.2010, red).
  Green dashed line denotes the considered LIS (Equation~\ref{Eq:LIS}).}
\label{Fig:fit}
\end{figure}

In order to describe the GCR differential energy spectrum near Earth we employed the widely used
 force-field approximation \citep[e.g.][]{vainio09}:
\begin{linenomath*}
\begin{equation}\label{Eq:FF}
J_{i}(T)=J_{{\rm LIS}_{i}}(T+\Phi_{i})\frac{T^{2}-T^{2}_{r}}{(T+\Phi_{i})^{2}-T^{2}_{r}}
\end{equation}
\end{linenomath*}
where $J_{{\rm LIS}_{i}}$ is the LIS, $T_r=0.938$ GeV is the proton's rest mass,
 $\Phi_i$ is the mean energy loss of the GCR particle inside the heliosphere, as defined by the modulation potential $\phi$:
 $\Phi_i=\phi\cdot(eZ_{i}/A_{i})$, where $Z_i$ and $A_i$ are the charge and mass numbers of the nucleus of type $i$.
Here we use the LIS from \citet{vos15}, which is a new parameterization based on both near-Earth and
 distant (beyond the heliospheric termination shock) measurements of GCR spectra and compositions.
The LIS takes the following form for protons:
\begin{equation}
J_{\rm LIS} = 2.7\cdot 10^3\, \,{T^{1.12}\over \beta^2}\left({T+0.67 \over 1.67}\right)^{-3.93},
\label{Eq:LIS}
\end{equation}
where $\beta=v/c$ is the ratio of the proton's velocity to the speed of light, $J$ and $T$ are given in
 [m$^2$ sec sr GeV/nuc]$^{-1}$ and GeV/nucleon, respectively.
This LIS is shown in Fig.~\ref{Fig:fit}.
We note that there are some other recent LIS estimates \citep[e.g.,][]{potgieter14,cummings16} which differ
 from each other mostly in the low energy part.
In order to account for that, \citet{corti16} proposed an additional parameter describing modulation for
 GCR protons with energy below 125 MeV, to which NMs are however insensitive.

It is important to consider $\alpha-$particles (effectively including heavier species) separately from protons
 since they are modulated differently and contribute 30--50\% to the overall count rate of a NM \citep{usoskin_bazi_11, caballero12}.
For $\alpha-$particles (including the heavier species) we used the same form as for protons (Eq.~\ref{Eq:LIS}) but with
 the weight of 0.3 (in the number of nucleons) similarly to \citet{usoskin_bazi_11}.
The intensity in this case is given for nucleons, and kinetic energy in GeV/nuc.
%


\section{PAMELA data}
\label{Sec:PAMELA}

The data used here include direct measurements of GCR energy spectra by the PAMELA space mission
 \citep{adriani11}, which is a space-borne magnetic spectrometer installed onboard the low orbiting
 satellite Resurs-DK1 with a quasi-polar (inclination $70^\circ$) elliptical orbit (height 350--600 km).
PAMELA was in operation since Summer 2006 through January 2016 continuously measuring all charged
 energetic ($> 80$ MeV) particles in space.

Here we make use of PAMELA the measurements of the differential energy spectra of CR protons obtained between
 July 2006 and January 2010, during which time the solar activity varied between moderate and very low.
This period was divided into 47 unequal time intervals, and the measured proton energy spectrum were
 provided by \citet{adriani13} (digital data are available at http://tools.asdc.asi.it/cosmicRays.jsp?tabId=0).
The month of December 2006 was excluded from consideration because of large disturbances of the CR flux
 due to a major Forbush decrease and a ground level enhancement \#70 \citep{usoskin_PAMELA_15}.

\begin{figure}[t]
\centering
\includegraphics[width=\columnwidth]{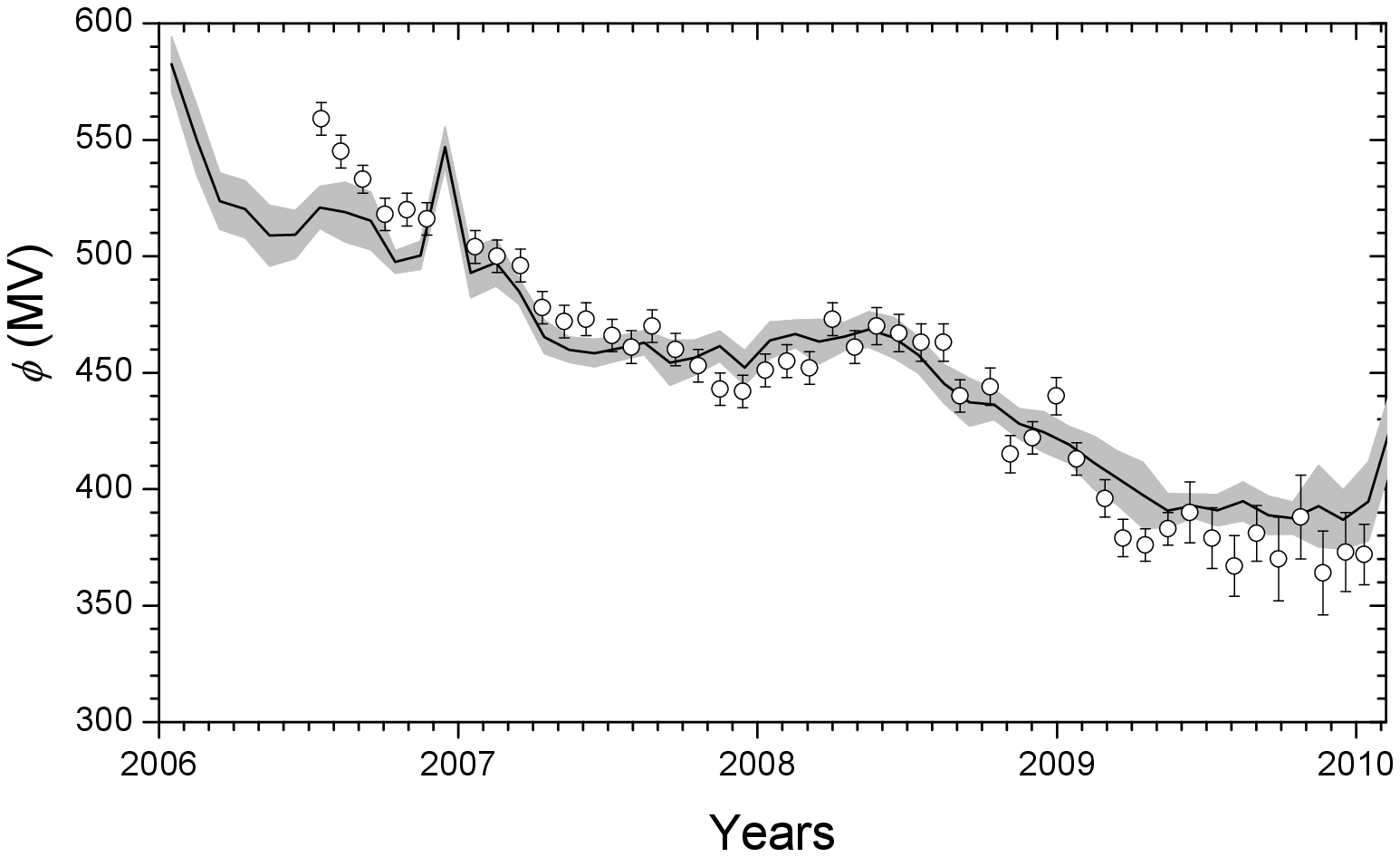}
\caption{The modulation potential $\phi$, along with its $1\sigma$ uncertainties, during 2006--2009, as obtained from PAMELA
 measurements (dots with error bars) as well as reconstructed from NM data (black line with grey shading).
 The spike in the curve corresponds to December 2006.}
\label{Fig:PAMELA}
\end{figure}

Each measured proton spectrum was fitted with the force-field model (Eqs.~\ref{Eq:FF} -- \ref{Eq:LIS}).
The best-fit value of $\phi$ and its $1\sigma$ uncertainties were found for each time interval
 by minimizing $\chi^2-$statistics in the interval of energies between 1--30 GeV, which
 corresponds to the most effective energy of GCR detection by NMs.
Two examples of the fit are shown in Fig.~\ref{Fig:fit} for time intervals S1 and S47.
The obtained values of the modulation potential are shown in Fig.~\ref{Fig:PAMELA}.

\section{Selection and calibration of NMs}
\label{Sec:calib}

Although the formalism (Section~\ref{Sec:model}) provides a full theoretical basis to the model count rate of an
 ideal standard NM, real instruments are neither ``ideal'' nor perfectly standard: different surrounding structures,
 instrumental setups (e.g., the electronic dead-time, high voltage, number of counters, the material for moderator, etc.),
 type of the counter (Soviet/Russian analogs CNM-15 are about 15\% less effective than the standard BP28(NM64) counters,
 \citep{gil15}), etc., making their sensitivities slightly different from each other.
Another source of the difference is that the reference barometric pressure can be set differently for different NMs,
 which can also result in the count rate being systematically deviating from the modeled one.
One approach to deal with that is to perform direct Monte-Carlo simulation of every NM considering the detailed
 geometry and environment \citep[e.g.,][]{aiemsa15,mangeard16}.
However, it is hardly possible to perform such detailed simulations for all NMs.
Accordingly, we consider this uncertainty as a constant scaling factor, which is defined individually for
 each NM, as described below.
This procedure is called ``calibration'' here.

For the analysis we selected sea-level and low-altitude ($\le 200$ m) NMs with long operation period and high stability.
The list of the selected NMs is given in Table~\ref{Tab:NM} along with their parameters.
\begin{table*}
\caption{Parameters of the neutron monitors used in the calculations for the period of July-2006 through December-2009.
Columns are: name, geomagnetic vertical effective cutoff rigidity $P_c$, altitude $h$, geographical coordinates and type of the NM,
 years of operation, as well as the scaling factor $\kappa$ (see text), and data source, respectively.
}\label{Tab:NM}
\begin{tabular}{p{1.5cm} c c c c c c p{3cm}}
\hline
 NM   & $P_c$ [GV] & $h$ [m] & Coordinates & type & Years & scaling $\kappa$ & Data source\\
\hline
  Moscow&2.43&200&37.32E 55.47N &24-NM64 & 04/1966 -- 05/2016 &$1.380$& NMDB$^a$, IZMIRAN$^b$\\
  Newark&2.4&50&75.75W 39.68N &9-NM64 & 07/1964 -- 05/2016 & $1.223$& NMDB \\
  Kerguelen&1.14&33&70.25E 49.35S &18-NM-64 & 02/1964 -- 01/2016 &$1.078$ & NMDB \\
  Oulu&0.8&15&25.47E 65.05N &9-NM6 & 04/1964 -- 05/2016 &$1.121$ & cosmicrays.oulu.fi\\
  Apatity$^\dagger$&0.65&181&33.4E 67.57N &18-NM-64 & 05/1969 -- 12/2015 &$1.869$ & pgia.ru/CosmicRay/\\
  Inuvik&0.3&21&133.72W 68.36N &18-NM-64 & 07/1964 -- 05/2016 & $1.254$& NMDB since 2000, IZMIRAN before\\
  McMurdo$^*$&0.3&48&166.6E 77.9S &18-NM-64 & 02/1964 -- 05/2016 &$0.875$& NMDB\\
  Kiel$^*$&2.36&54&10.12E 54.34N &18-NM-64 & 09/1964 -- 12/2014 &$1.395$& NMDB \\
\hline
\end{tabular}
$^\dagger$Long dead-time.\\
$^*$not used in the final reconstruction (see Section~\ref{Sec:test}).\\
$^a$http://www.nmdb.eu/\\
$^b$http://cr0.izmiran.ru/common/links.htm
\end{table*}

\begin{figure}[t]
\centering
\includegraphics[width=\columnwidth]{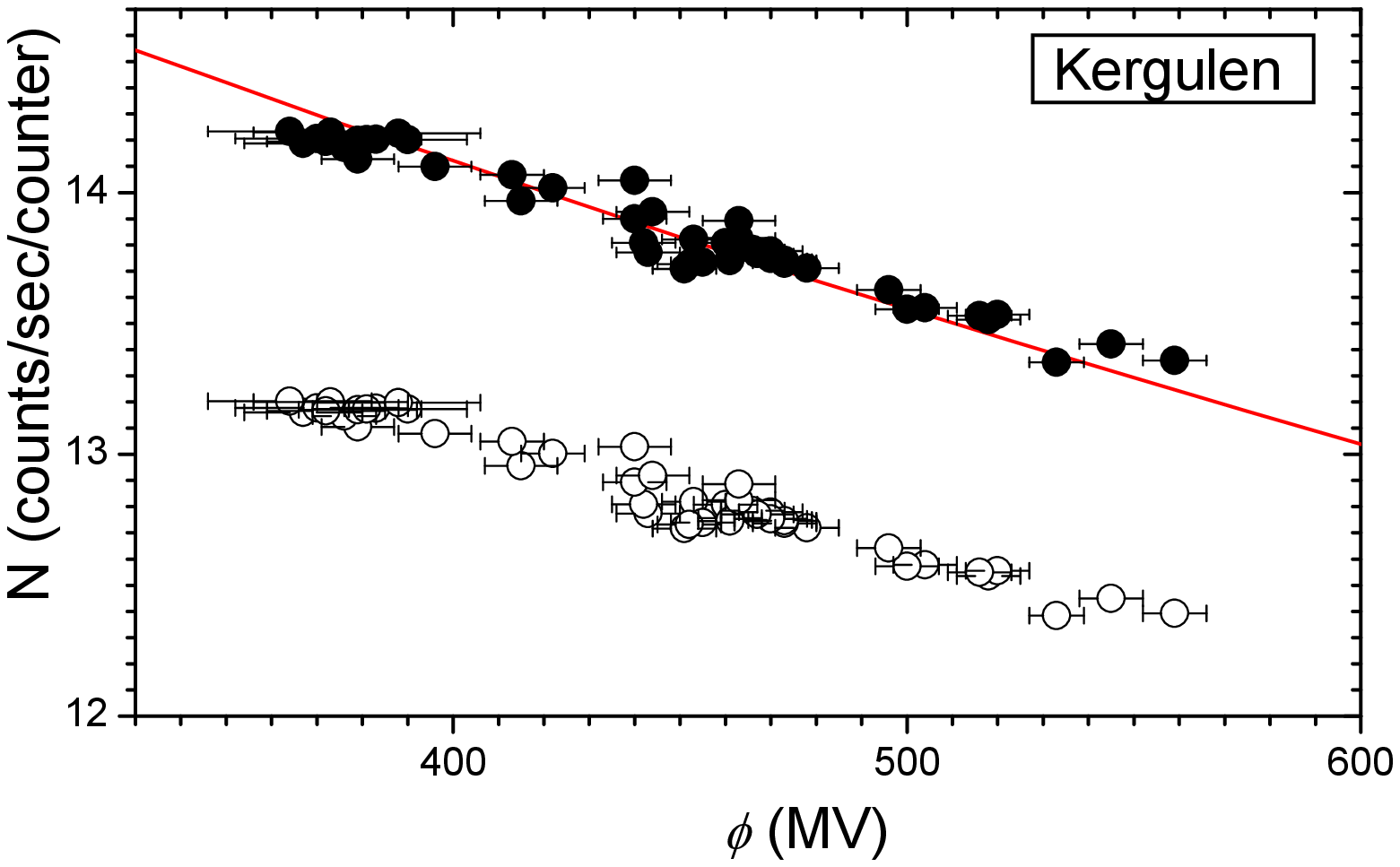}
\caption{The mean count rate of the Kergulen NM during periods of PAMELA measurements (the corresponding values of $\phi$ are
 shown as the X-values) as recorded (open dots) and scaled using the best-fit factor $\kappa=1.078$.
 The red curve represents the model.}
\label{Fig:kerg}
\end{figure}

Using the best-fit values of $\phi_i$ (with uncertainties) obtained for 47 PAMELA spectra (Section~\ref{Sec:PAMELA}), we
 calculated, using Eqs.~\ref{Eq:N}--\ref{Eq:LIS}, the expected count rates of a standard ideal NM for the same periods
 when PAMELA measured spectra, $N^*_i$.
For the same 47 periods we collected the actual mean count rates, $N_i$ for each NM.
Then the scaling factor $\kappa_i = N^*_i/N_i$ was calculated with its uncertainties.
Finally, from 47 values of $\kappa_i$ we defined, using the standard weighted averaging,
 the mean scaling factor $\kappa$ for each NM, as shown in Table~\ref{Tab:NM}.
The formal standard error of the mean $\kappa$ is small (0.001--0.002) and is not shown.
The fact that the errors are small for different modulation levels implies that indeed
 the method works, and the scaling factor $\kappa$ adequately described the non-ideality of a NM.
An example is shown in Fig.~\ref{Fig:kerg} for the Kergulen NM.
While the recorded count rates (open dots) lie systematically below the model curve, implying that
 this NM is slightly less effective than the ``ideal'' standard one, the use of the best-fit scaling
 factor $\kappa=1.078$ makes the data fully consistent with the model curve.

\section{Reconstruction of the modulation potential}
\label{Sec:recon}

\begin{figure*}[t]
\centering
\includegraphics[width=\textwidth]{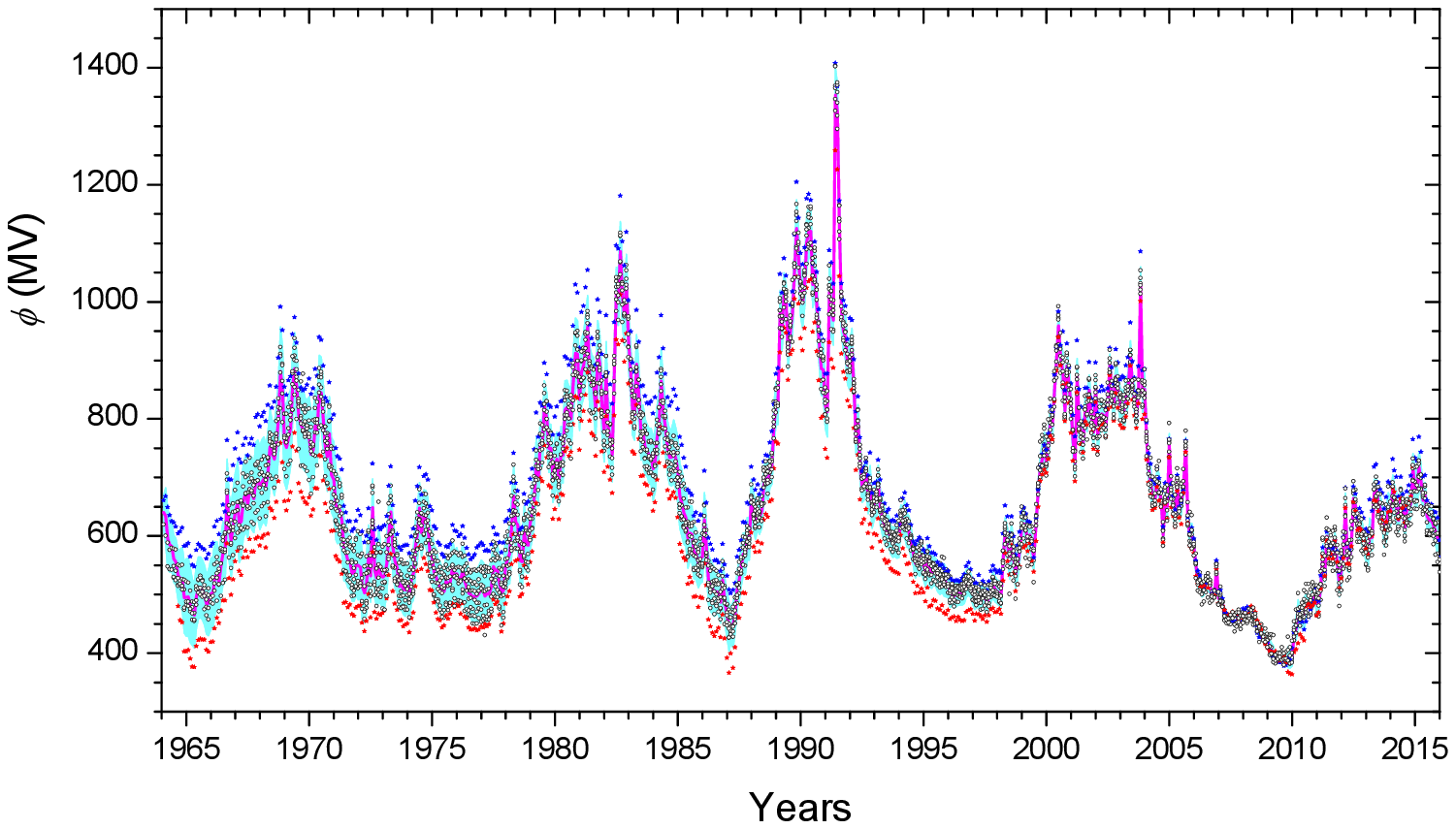}
\caption{Reconstruction of the monthly modulation potential values (small dots) from individual NMs listed in Table~\ref{Tab:NM}.
 The red and blue dots correspond to Kiel and McMurdo NMs, respectively.
 The magenta line and the light blue shading represent the mean and the standard deviation of these individual reconstructions.}
\label{Fig:series}
\end{figure*}

Once the scaling factor $\kappa$ is fixed for a given NM, the problem can be inverted so that from the
 measured (and corrected using the factor $\kappa$) count rate one can calculate the corresponding value of
 the modulation potential $\phi$.
We did it by calculating the monthly values of $\phi$ for each NM listed in Table~\ref{Tab:NM}.
The result is shown in Figure~\ref{Fig:series} with small dots.
One can see that the spread of dots is very small during and around the calibration period in 2006--2010,
 but they diverge in the earlier part of the period, in the 1960--1970s.

In the analysis, we considered also slow changes in the geomagnetic cutoff rigidity for each NM.

\subsection{Long-term consistency of the NMs}
\label{Sec:test}

\begin{figure*}[t]
\centering
\includegraphics[width=\textwidth]{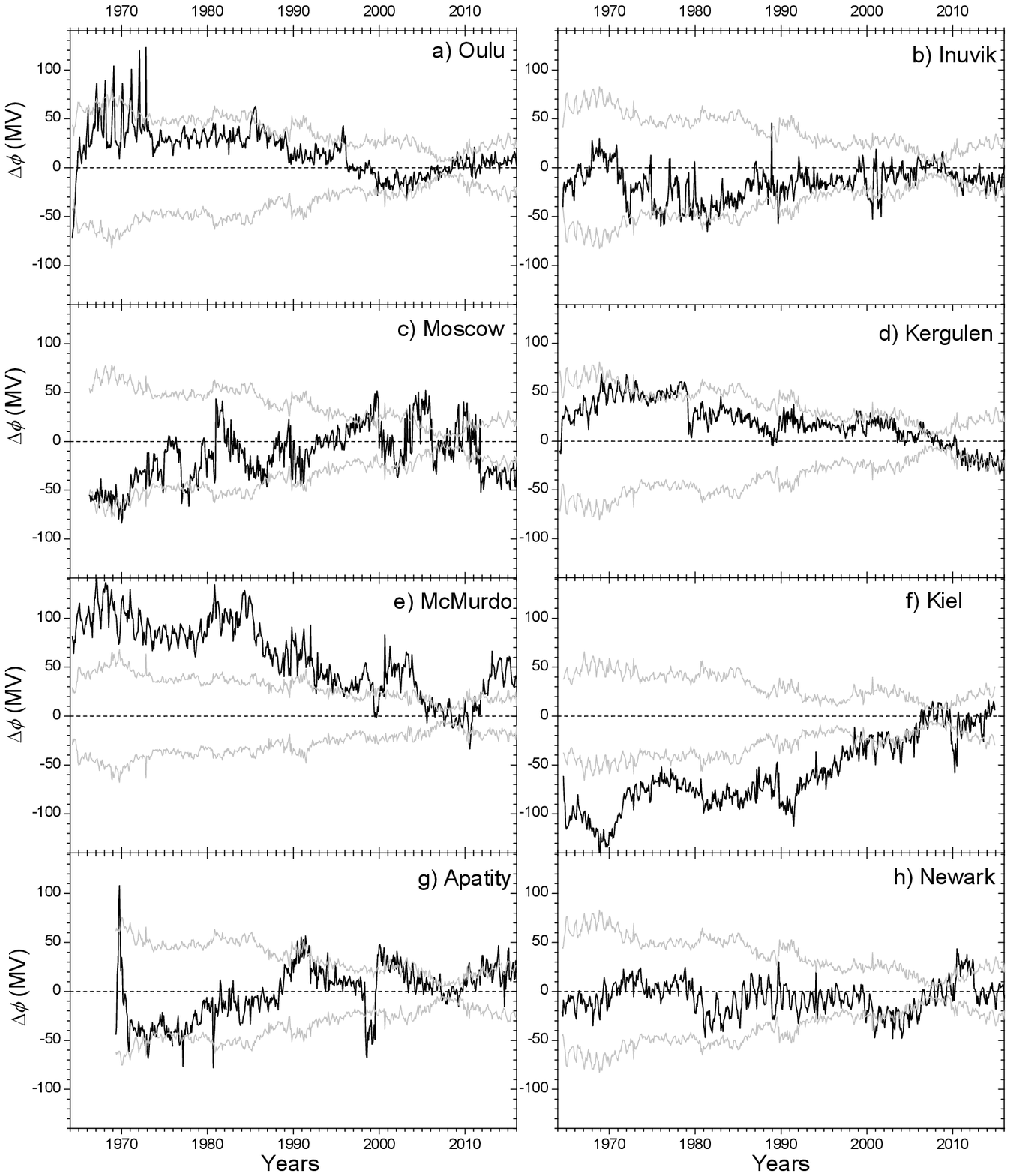}
\caption{The difference $\Delta\phi$ between values of the modulation potential, computed from individual NMs (as indicated in the
 legends), and the mean $\phi$ computed using data from all the available NMs but this one.
 The grey lines bound the $\pm 1\sigma$ uncertainties of the mean series.}
\label{Fig:compare}
\end{figure*}

Next, we check each of the analyzed NMs for long-term consistency.
For that, we calculated the difference $\Delta\phi$ between the modulation potential calculated only from this NM data
 and that, calculated (as the mean and standard deviation) from the data of the other seven NMs in Table~\ref{Tab:NM}, excluding the tested one.
This is shown in Figure~\ref{Fig:compare}.

We note that the \textit{Oulu NM} (Fig.~\ref{Fig:compare}a) exhibits small ($\pm 25$ MV) deviations from the mean curve, but is always
 within the $1\sigma$ uncertainty of the latter, except for strong seasonal peaks in the earlier part.
These peaks were caused by snow on the roof during winter months before 1974, when the Oulu NM was finally settled in a building with
 a pyramid-shaped warmed roof, so that snow was never accumulated above the NM since 1974.
To avoid the uncontrolled effect of snow, we have excluded from further consideration the Oulu NM data for months January through March
 for years 1964--1973.
The first months of data, in 1964, also depict a strong drift and were removed.

\textit{Inuvik, Moscow and Kergulen NMs} (Fig.~\ref{Fig:compare}b--d) exhibit deviations up to $\pm 50$ MV from the mean curve, but
 are mostly within the $1\sigma$ uncertainty of the latter.
Accordingly, data from these NMs were considered as they are.

The \textit{McMurdo NM} (Fig.~\ref{Fig:compare}e) depicts a strong systematic deviation greatly exceeding $1\sigma$ limits,
 that was as much as about 100 MV before the 1980s.
This is clearly seen in Fig.~\ref{Fig:series}, where the blue dots lie systematically above the curve.
This trend implies that the McMurdo NM tends to increase its count rate in time against other stations.

On the contrary, the \textit{Kiel NM} (Fig.~\ref{Fig:compare}f) depicts an opposite but equally strong
 trend in the deviation, also exceeding systematically the $1\sigma$ limit.
This is seen in Fig.\ref{Fig:series} as a systematic divergence of the red points.
The systematic growth of $\Delta\phi$  implies that the Kiel NM count rate decreases in time against all other NMs.
Interestingly, these two NMs nearly compensate each other in the composite, but grossly increase the error bars.
Because of the systematic drifts, we do not include McMurdo and Kiel NMs into the final reconstruction of $\phi$.

The \textit{Apatity NM} (Fig.~\ref{Fig:compare}g) depicts deviations within $\pm 50$ MV from the mean curve, mostly within
 the $1\sigma$ uncertainty of the latter.
There is a spike in $\Delta\phi$ in 1969 (the first year of the NM operation).
Because of it we only use the Apatity NM data after 1970.
There is another spike in 1998--1999, but no correction was implied for this.

The \textit{Newark NM} (Fig.~\ref{Fig:compare}g) depicts some deviations within $\pm 30$ MV from the mean curve, mostly always
 within the $1\sigma$ uncertainty of the latter.
It also depicts a seasonal cycle, but it is small and not a subject to correction or removal.

Thus, from the eight preliminary selected NMs we use for further analysis six most stable ones
 (Oulu, Inuvik, Moscow, Kergulen, Apatity and Newark), while McMurdo and Kiel NMs depict systematic
 drifts and are not considered henceforth.

\subsection{Extension before 1965}
\label{Sec:1965}

\begin{figure}[t]
\centering
\includegraphics[width=\columnwidth]{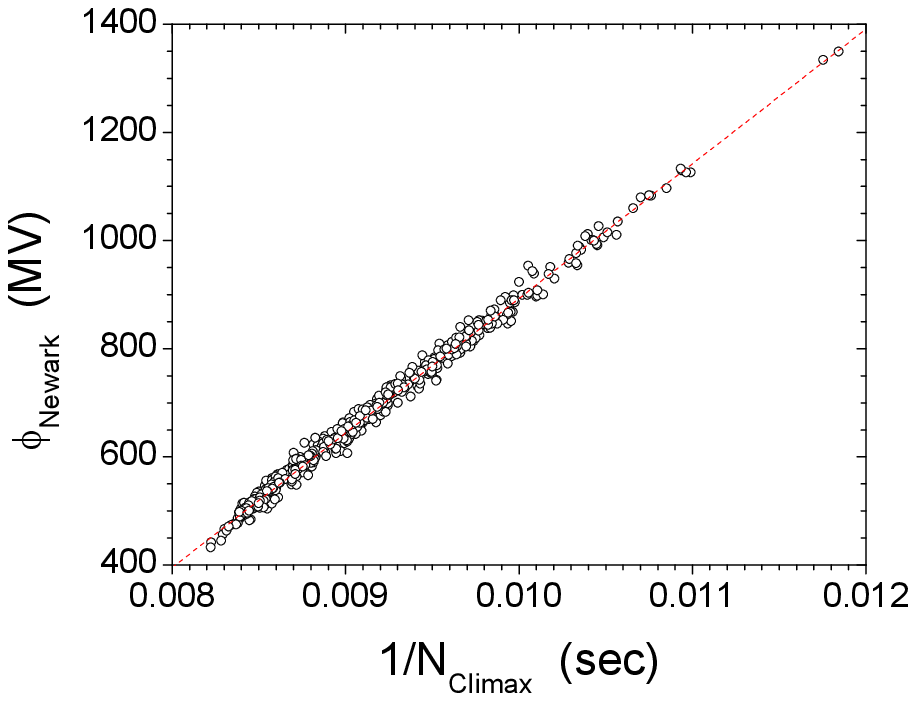}
\caption{Scatter plot of monthly $\phi$ reconstructed from the Newark NM data for the period 1965--2000 and the inverted count rate
 of the Climax NM for the same period.
 The best-fit linear regression is shown by the red dotted line.}
\label{Fig:Climax}
\end{figure}

The consideration above was based on the World network of neutron monitors of the standard type NM64, which
 was introduced in 1964.
Before that, there were several NMs of another design, called IGY (International Geophysical Year), the longest record
 being from the Climax NM (altitude $\approx 3400$ m, $P_c\approx$3 GV) from February 1951 until November 2006, and Mt.Washington
 ($\approx 1900$ m, $P_c\approx$ 1.3 GV) from November 1955 until June 1991.
However, since they were not in operation during the PAMELA calibration period, their calibration was done via the
 overlap with the main world NM network since 1964.
Because of their mid- and high-altitude location, the theoretical model (available for the sea level) is not applied.
Accordingly, following the approach of \citet{usoskin_bazi_11}, for these two NMs we used an empirical relation
 between the NM count rate $N$ and the modulation potential $\phi$:
\begin{equation}
\phi = {A\over N} + B,
\label{Eq:Climax}
\end{equation}
where $A$ and $B$ are free parameters.
Figure~\ref{Fig:Climax} shows a scatter plot of the monthly values of $\phi$ reconstructed from
 the Newark NM for the period 1965--2006 vs. the Climax NM's inverted count rate.
One can see that the relation (Eq.~\ref{Eq:Climax}) is nearly perfectly linear and can be fitted (using the standard linear least-square method)
 with $A=(2.49\pm 0.02)\cdot 10^5$ MV/sec and $B= 1597\pm 18$ MV.
This Newark-vs-Climax relation can be used to estimate $\phi$ before 1965 from the Climax NM data.
We constructed similar relations for all the six selected NMs, and thus have six series of $\phi$ for the period 1951--1964.

A similar analysis was performed also for the Mt.Washington NM data since 1955.
As a result, for each month for the period 1955--1964 we have 12 values of $\phi$ (6 from Climax and 6 from Mt.Washington), from
 which we calculated the mean and the standard deviation as an assessment of the modulation potential for that period.
For the period 1951--1955 only six $\phi$ series were used.

\section{Results and discussions}
\label{Sec:result}
\subsection{Final series}

\begin{figure}[t]
\centering
\includegraphics[width=\columnwidth]{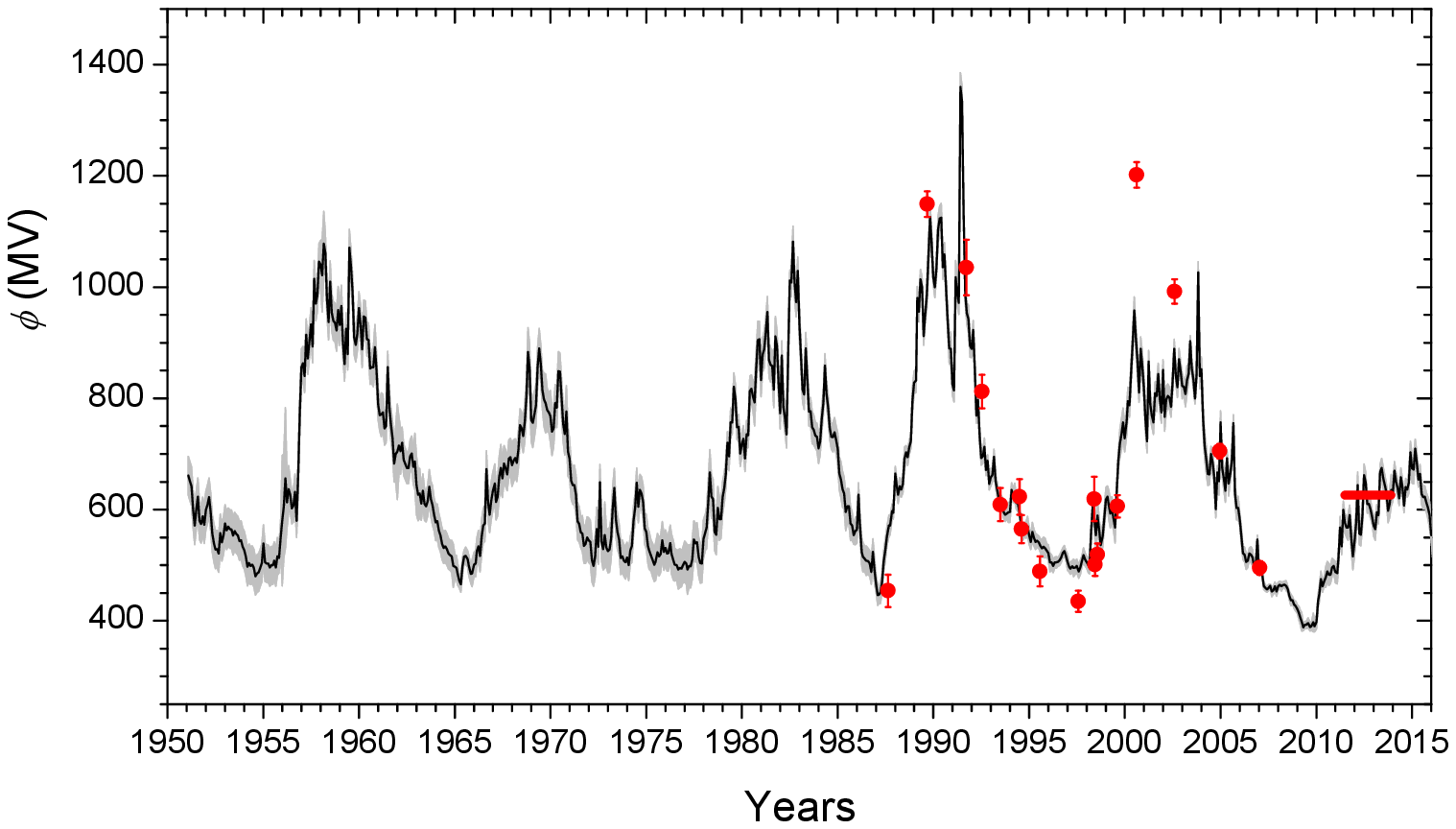}
\caption{Monthly series of the modulation potential reconstructed here (black curve with grey shading depicting the $1\sigma$ uncertainties)
 along with the values of $\phi$ obtained by fitting balloon data.
 The data series is available in Table~\ref{Tab:mon}.
 The red stripe represents $\phi$ obtained from a fit to AMS-02 data.}
\label{Fig:balloon}
\end{figure}

The final reconstruction of the modulation potential is shown in Figure~\ref{Fig:balloon} along with its $1\sigma$ uncertainties,
 while digital values are given in Table~\ref{Tab:mon}.

\begin{figure}[t]
\centering
\includegraphics[width=\columnwidth]{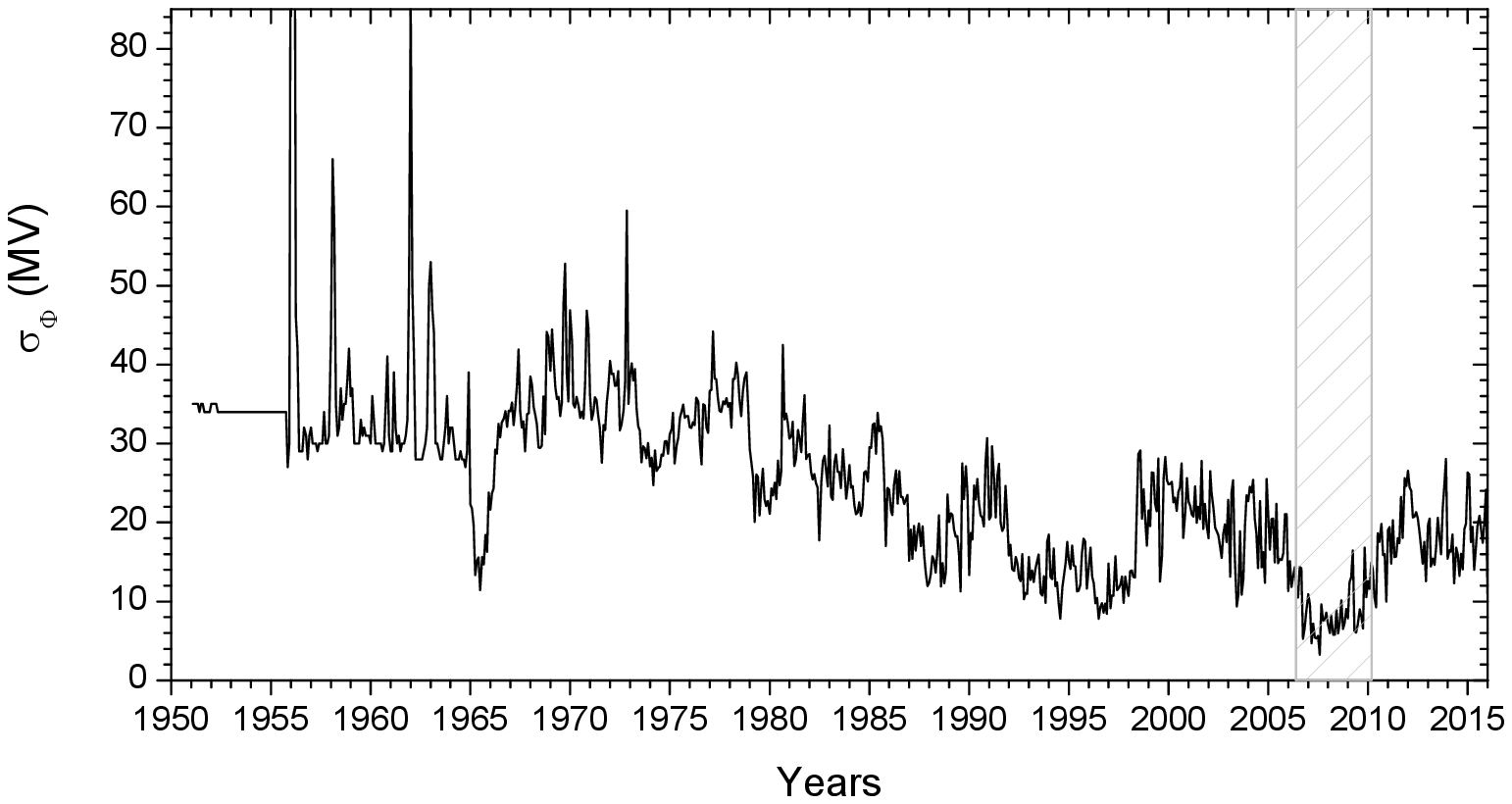}
\caption{Standard error of the monthly modulation potential reconstruction (see Figure~\ref{Fig:series}).
The grey shaded region denotes the period of PAMELA calibration.}
\label{Fig:error}
\end{figure}

The uncertainties are also shown separately in Figure~\ref{Fig:error}.
One can see that the uncertainty is small ($<10$ MV) during the PAMELA calibration period (denoted by the grey shading),
 moderate (20--25 MV) after the 1980's and gradually increases back in time reaching $\approx 35$ MV around 1970,
 and stays at roughly this level before that.
For the period 1955--1964, the spikes are caused by a discrepancy between Climax and Mt.Washington data.
Before 1955, only the Climax NM data is available and the uncertainty is flat.


\subsection{Comparison with other direct measurements}

Beyond the PAMELA data, used for calibration of the NMs, we now compare the final modulation potential series
 with the values of $\phi$ obtained by fitting GCR spectra from short-time space- and balloon-borne measurements
 as shown in Figure~\ref{Fig:balloon}.
We used the following balloon- and space-borne data (see full details and data collection at
 http://tools.asdc.asi.it/cosmicRays.jsp?tabId=0) for the following measurements
 of the GCR energy spectrum: LEAP, MASS89, MASS91, IMAX92, POLAR, POLAR-2, BESS-TeV, BESS00, BESS93, BESS94,
 BESS95, BESS97, BESS98, BESS99, CAPRICE, CAPRICE98, AMS-01, AMS-02, with the original references to
 \citep{seo91,seo01,webber91,bellotti99,boezio99,boezio03,alcaraz00,menn00,wang02,shikaze07,adriani13,aguilar15,abe16}

One can see a general agreement between the overall curve and the individual points excepts for two balloon points,
 BESS00 and BESS-TeV, yielding too strong modulation in 2000 and 2002, respectively, and one point, BESS97,
 implying too low modulation in 1997.
Note that disagreement of these data points with the NM data was mentioned also by \citet{ghelfi16}.

\begin{figure}[t]
\centering
\includegraphics[width=\columnwidth]{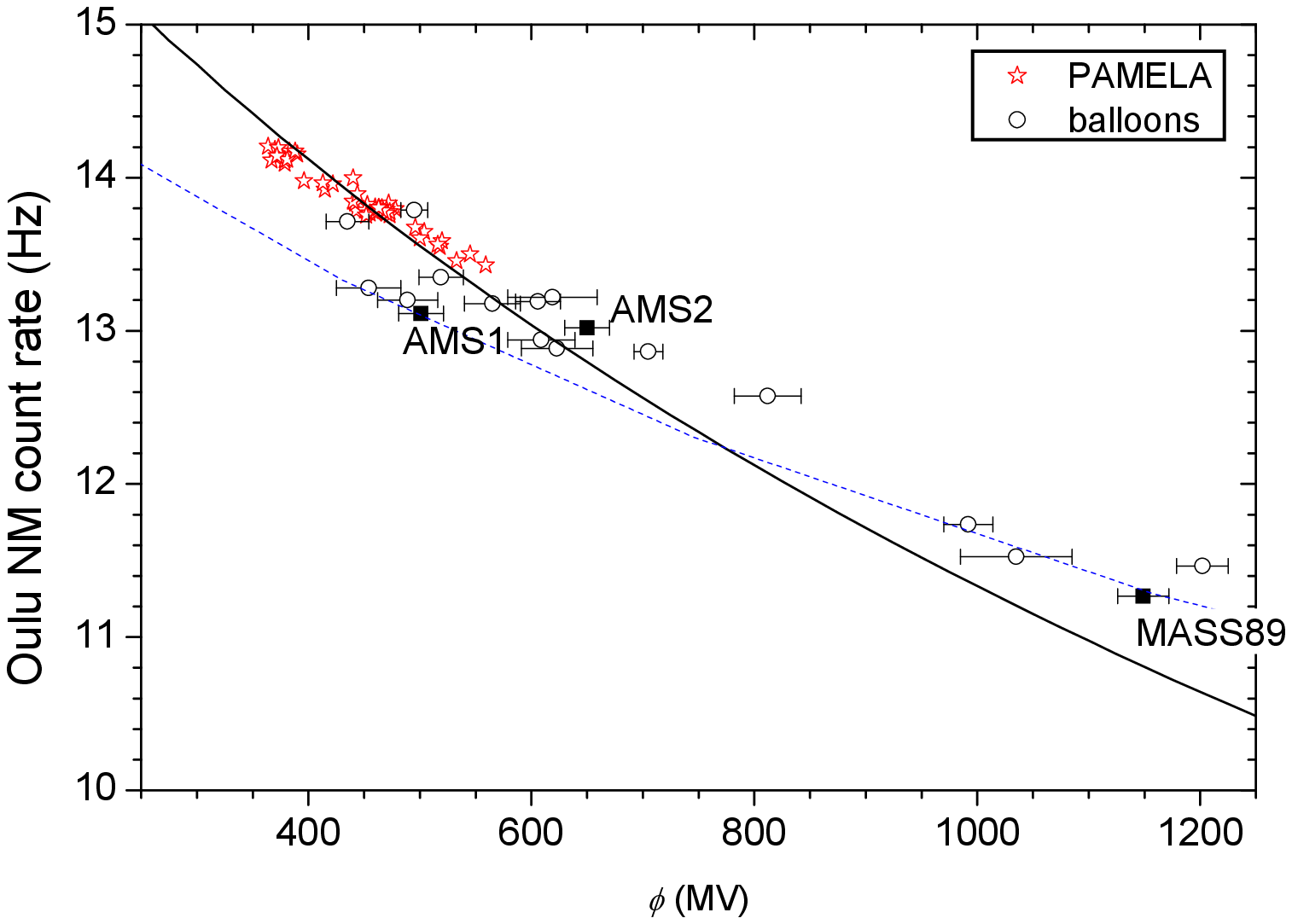}
\caption{Comparison of this model (black solid curve) and experimental (points) count rates of the Oulu NM for the periods
 when measurements of the GCR spectra were available.
 Red stars -- PAMELA periods used for calibration here;
 open circles -- different balloon measurements;
 black squares -- AMS-01 and MASS89, used for calibration in \citet{usoskin_Phi_05} and \citet{usoskin_bazi_11}, as well as AMS-02.
 The dotted blue lines shows the model curve from \citet{usoskin_bazi_11}.}
\label{Fig:scatter}
\end{figure}

However, such a comparison is not very representative, since the reconstructed series is with monthly resolution while
 individual flights had duration from several hours to several days, or, as in the case of AMS-02 data taking, several years.
In Figure~\ref{Fig:scatter} we show a scatter plot of the Oulu NM count rates (scaled with the factor 1.121 -- see Table~\ref{Tab:NM},
 statistical errors are negligible) averaged over exactly the same periods as data taking for the balloon- and space-flight,
 vs. the fitted values of $\phi$ for these flights (as shown by red dots in Fig.~\ref{Fig:balloon}).
The solid black curve shows the model-predicted dependence between a polar NM count rate and the modulation potential.
One can see that the agreement is quite good for the range of $\phi-$values covering PAMELA data
 used for calibration (red stars), viz. 300--700 MV.
While PAMELA data lie tightly along the model curve, other data produce a large scatter, but still
 around the curve.
On the other hand, data points lie slightly but systematically above the curve in the range of higher $\phi-$values
 800--1200 MV.
In particular, BESS00 and BESS-TeV, mentioned above, and MASS89 (used for calibration by \citet{usoskin_bazi_11}) balloon data
 suggest a higher modulation than the model does, during periods of the active Sun.
This may indicate that the model may slightly underestimate the modulation during such periods,
 but the lack of reliable data in this range (only 4--5 points vs. $\approx 60$ in the lower activity range) does
 not provide a solid ground for such a conclusion.
The relations for other NMs (not shown) are similar to this example.

Thus, we conclude that the new reconstruction of the modulation potential is in good agreement with fragmentary
 direct measurements, at least for the periods of low and moderate solar activity.

\subsection{Comparison with the previous reconstructions}

We emphasize that the values of the modulation potential presented here, should not be directly
 compared with those published earlier.
The reason is that the value of $\phi$ has no absolute physical meaning and depends on the LIS models
 used in its calculation \citep{usoskin_Phi_05,herbst10}.
It is only a useful parameter to describe the energy spectrum of GCR near Earth, with the fixed LIS value.
Therefore, the fact that the values of $\phi$ presented here are different from the earlier ones does not
 imply different GCR spectra.

\begin{figure}[t]
\centering
\includegraphics[width=\columnwidth]{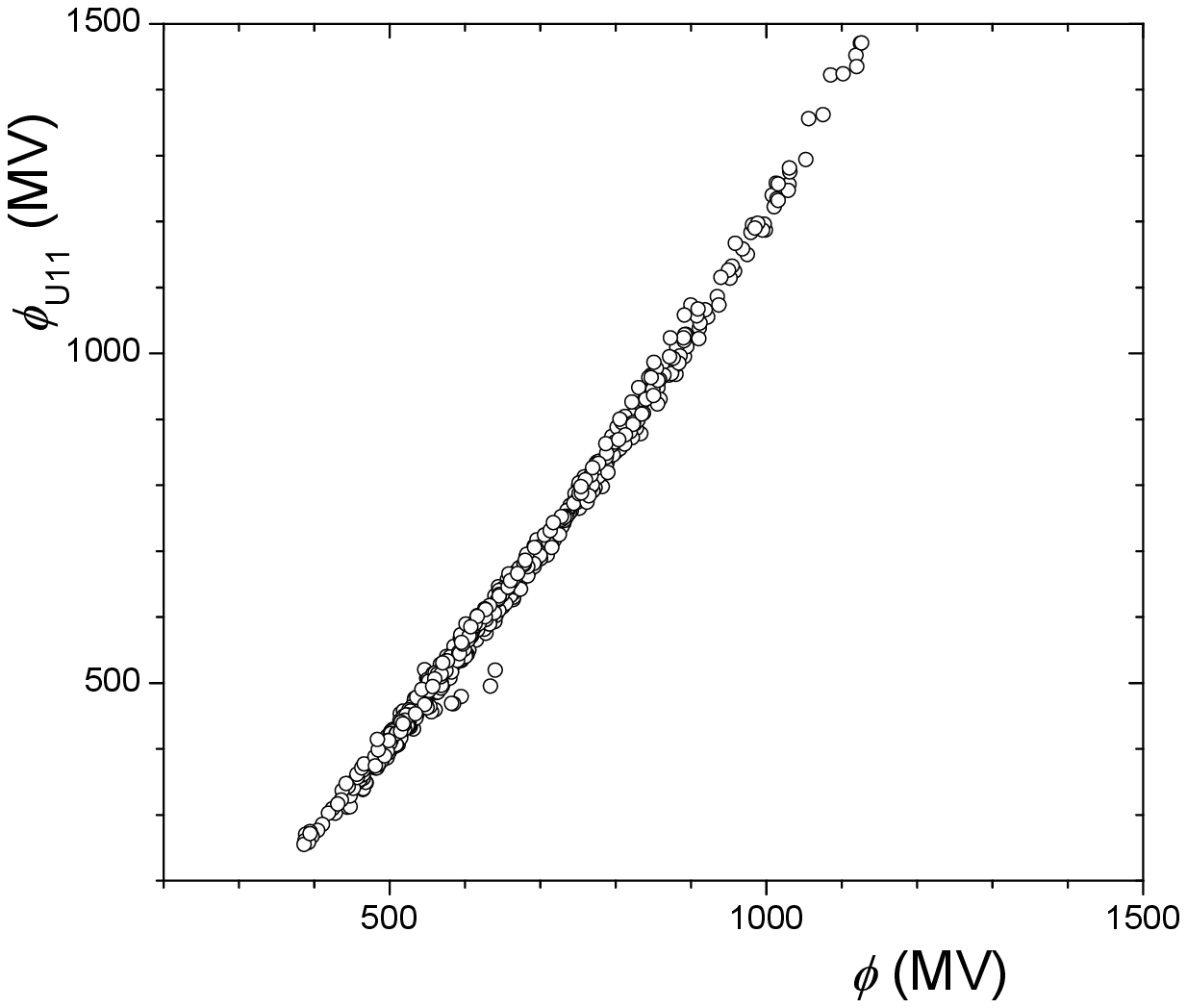}
\caption{Comparison of monthly modulation potentials obtained here (X-axis) and those from \citep{usoskin_bazi_11} (Y-axis).}
\label{Fig:fi-fi}
\end{figure}

A scatter plot of the previously published modulation potential series \citep{usoskin_bazi_11} versus the
 results of this work is shown in Figure~\ref{Fig:fi-fi}.
The relation is very tight and slightly nonlinear.

The difference from the earlier models is caused by three main facts:
(1) the use of the new NM yield function \citep[][see also erratum therein]{mishev13};
(2) the calibration of the NM responses directly to a large set of PAMELA data, while earlier models were
 linked to two points -- AMS-01 and MASS89 (see Figure~\ref{Fig:scatter}); and
(3) the use of the updated LIS \citep{vos15}.

\section{Conclusions}
\label{Sec:conc}

We have presented a new reconstruction of the heliospheric modulation potential for galactic cosmic rays
 during the neutron monitor era, since 1951.
The new reconstruction is based on a new-generation specific yield function of a NM, exploits an updated model
 of the LIS, and applies a calibration to direct measurements of the GCR energy spectrum during 47 episodes by PAMELA
 space-borne spectrometer.
The reconstruction is based on data from six standard NM64-type neutron monitors (Apatity, Inuvik, Kergulen, Moscow, Newark and Oulu)
 since 1965, and two IGY-type NMs (Climax and Mt.Washington) before that, all demonstrating stable operation over the decades.
The new reconstruction is presented in Table~\ref{Tab:mon}.

We also tested the long-term stability of individual NMs and found that McMurdo and Kiel NMs exhibit essential
 drifts, while all other analyzed NMs are fairly stable on the multi-decadal time scale.

The presented series forms a benchmark record of the cosmic ray variability (in the NM energy range) for the last 60 years,
 and can be used to long-term studies in the fields of solar-terrestrial physics, atmospheric sciences, etc.

\begin{table*}
\caption{Mean monthly value of the modulation potential (see Figure~\ref{Fig:balloon}) for 1951--2016.}
\label{Tab:mon}
\begin{tabular}{c|cccccccccccc}
\hline
Yr/month	&	Jan	&	Feb	&	Mar	&	Apr	&	May	&	Jun	&	Jul	&	Aug	&	Sep	&	Oct	&	Nov	&	Dec	\\
\hline
1951	&	N/A	&	661	&	651	&	642	&	601	&	571	&	600	&	623	&	580	&	574	&	588	&	573	\\
1952	&	599	&	609	&	622	&	595	&	557	&	540	&	528	&	529	&	521	&	558	&	541	&	553	\\
1953	&	575	&	563	&	569	&	565	&	564	&	553	&	561	&	556	&	553	&	543	&	545	&	533	\\
1954	&	527	&	515	&	497	&	504	&	499	&	501	&	495	&	480	&	485	&	487	&	495	&	504	\\
1955	&	539	&	505	&	503	&	503	&	496	&	500	&	500	&	508	&	496	&	516	&	513	&	549	\\
1956	&	580	&	606	&	656	&	613	&	637	&	627	&	602	&	607	&	632	&	580	&	663	&	775	\\
1957	&	856	&	863	&	840	&	914	&	872	&	902	&	933	&	893	&	1015	&	970	&	986	&	1046	\\
1958	&	1040	&	1021	&	1078	&	1061	&	978	&	937	&	1010	&	954	&	939	&	936	&	922	&	959	\\
1959	&	932	&	966	&	900	&	862	&	925	&	880	&	1071	&	1038	&	992	&	909	&	896	&	916	\\
1960	&	962	&	932	&	888	&	947	&	945	&	906	&	905	&	854	&	856	&	856	&	892	&	850	\\
1961	&	789	&	770	&	771	&	775	&	746	&	750	&	856	&	784	&	753	&	729	&	683	&	700	\\
1962	&	704	&	715	&	705	&	720	&	689	&	682	&	675	&	675	&	696	&	701	&	677	&	686	\\
1963	&	644	&	627	&	629	&	611	&	634	&	610	&	606	&	615	&	641	&	619	&	608	&	591	\\
1964	&	577	&	579	&	562	&	553	&	538	&	531	&	534	&	529	&	521	&	518	&	516	&	499	\\
1965	&	497	&	495	&	484	&	472	&	466	&	498	&	515	&	517	&	511	&	499	&	484	&	485	\\
1966	&	501	&	503	&	525	&	534	&	517	&	551	&	571	&	575	&	673	&	614	&	591	&	616	\\
1967	&	633	&	647	&	614	&	623	&	659	&	675	&	656	&	683	&	674	&	663	&	691	&	694	\\
1968	&	680	&	688	&	693	&	684	&	708	&	752	&	746	&	732	&	758	&	801	&	883	&	854	\\
1969	&	761	&	756	&	771	&	788	&	864	&	897	&	861	&	812	&	784	&	774	&	770	&	772	\\
1970	&	768	&	740	&	746	&	791	&	783	&	849	&	848	&	801	&	754	&	736	&	776	&	704	\\
1971	&	691	&	652	&	644	&	644	&	627	&	579	&	574	&	559	&	557	&	535	&	538	&	550	\\
1972	&	545	&	542	&	509	&	499	&	516	&	565	&	533	&	649	&	542	&	529	&	554	&	532	\\
1973	&	525	&	528	&	548	&	603	&	639	&	582	&	559	&	546	&	513	&	516	&	508	&	505	\\
1974	&	514	&	500	&	527	&	535	&	580	&	603	&	648	&	605	&	635	&	629	&	614	&	566	\\
1975	&	560	&	535	&	527	&	517	&	511	&	501	&	506	&	523	&	516	&	520	&	545	&	524	\\
1976	&	531	&	521	&	522	&	540	&	520	&	508	&	501	&	500	&	492	&	495	&	494	&	499	\\
1977	&	506	&	503	&	492	&	499	&	498	&	510	&	546	&	540	&	541	&	515	&	504	&	503	\\
1978	&	547	&	560	&	568	&	613	&	667	&	623	&	619	&	560	&	555	&	603	&	583	&	583	\\
1979	&	622	&	638	&	664	&	720	&	696	&	757	&	757	&	821	&	799	&	748	&	749	&	701	\\
1980	&	717	&	727	&	692	&	735	&	734	&	813	&	817	&	803	&	793	&	853	&	905	&	906	\\
1981	&	833	&	878	&	888	&	919	&	955	&	869	&	860	&	858	&	816	&	911	&	894	&	827	\\
1982	&	773	&	877	&	784	&	765	&	735	&	880	&	1012	&	1008	&	1082	&	1008	&	973	&	1029	\\
1983	&	936	&	876	&	816	&	807	&	890	&	834	&	776	&	777	&	748	&	743	&	733	&	731	\\
1984	&	710	&	722	&	763	&	791	&	859	&	807	&	784	&	747	&	731	&	730	&	739	&	725	\\
1985	&	709	&	665	&	654	&	637	&	629	&	596	&	606	&	606	&	575	&	571	&	554	&	561	\\
1986	&	564	&	627	&	578	&	524	&	515	&	511	&	508	&	507	&	503	&	488	&	524	&	491	\\
1987	&	466	&	446	&	447	&	451	&	470	&	507	&	527	&	551	&	571	&	571	&	597	&	594	\\
1988	&	665	&	640	&	627	&	642	&	636	&	646	&	691	&	703	&	694	&	712	&	723	&	796	\\
1989	&	828	&	832	&	980	&	956	&	1014	&	1000	&	912	&	952	&	986	&	1055	&	1126	&	1078	\\
1990	&	1018	&	1000	&	1034	&	1103	&	1123	&	1125	&	1036	&	1059	&	999	&	941	&	890	&	889	\\
1991	&	826	&	814	&	1018	&	990	&	972	&	1360	&	1334	&	1133	&	990	&	953	&	943	&	895	\\
1992	&	889	&	923	&	852	&	769	&	797	&	734	&	693	&	696	&	713	&	671	&	688	&	646	\\
1993	&	658	&	662	&	695	&	651	&	635	&	620	&	612	&	612	&	596	&	591	&	593	&	595	\\
1994	&	596	&	635	&	625	&	631	&	612	&	610	&	594	&	574	&	556	&	564	&	563	&	570	\\
1995	&	553	&	542	&	560	&	549	&	542	&	544	&	541	&	535	&	530	&	533	&	530	&	526	\\
1996	&	522	&	505	&	505	&	499	&	505	&	505	&	506	&	509	&	514	&	522	&	525	&	517	\\
1997	&	506	&	497	&	494	&	498	&	495	&	496	&	499	&	489	&	496	&	507	&	518	&	510	\\
1998	&	505	&	500	&	498	&	559	&	606	&	583	&	554	&	589	&	558	&	536	&	552	&	576	\\
1999	&	618	&	623	&	609	&	593	&	601	&	584	&	568	&	632	&	687	&	719	&	735	&	757	\\
2000	&	728	&	759	&	795	&	788	&	846	&	899	&	958	&	908	&	868	&	811	&	889	&	859	\\
2001	&	820	&	761	&	723	&	866	&	789	&	769	&	757	&	809	&	804	&	844	&	798	&	775	\\
2002	&	851	&	767	&	798	&	804	&	801	&	785	&	830	&	889	&	841	&	820	&	870	&	849	\\
2003	&	822	&	821	&	808	&	833	&	844	&	903	&	853	&	834	&	800	&	844	&	1026	&	840	\\
2004	&	852	&	766	&	715	&	691	&	664	&	663	&	700	&	685	&	660	&	601	&	667	&	651	\\
2005	&	757	&	675	&	656	&	634	&	692	&	647	&	663	&	686	&	755	&	629	&	603	&	603	\\
2006	&	582	&	552	&	522	&	519	&	506	&	508	&	521	&	519	&	514	&	496	&	500	&	546	\\
2007	&	492	&	498	&	484	&	463	&	459	&	456	&	459	&	463	&	453	&	455	&	462	&	453	\\
2008	&	462	&	465	&	462	&	464	&	465	&	463	&	457	&	445	&	437	&	437	&	429	&	426	\\
2009	&	421	&	414	&	407	&	398	&	389	&	393	&	393	&	396	&	389	&	390	&	397	&	390	\\
2010	&	398	&	436	&	448	&	475	&	462	&	471	&	478	&	489	&	483	&	483	&	496	&	499	\\
2011	&	488	&	486	&	513	&	568	&	534	&	599	&	576	&	568	&	568	&	590	&	549	&	516	\\
2012	&	542	&	574	&	644	&	557	&	555	&	590	&	662	&	649	&	610	&	610	&	601	&	587	\\
2013	&	571	&	565	&	595	&	590	&	667	&	675	&	656	&	644	&	630	&	598	&	610	&	635	\\
2014	&	630	&	670	&	650	&	639	&	620	&	658	&	640	&	607	&	640	&	636	&	648	&	703	\\
2015	&	674	&	672	&	709	&	684	&	654	&	664	&	628	&	622	&	623	&	612	&	603	&	590	\\
2016	&	554	&	526	&	531	&	529	&	524	&	525	&	537	&	519	&	518	&	497	&	482	&	483	\\
\hline
\end{tabular}
\end{table*}

\begin{acknowledgments}
Data of NMs count rates were obtained from http://cosmicrays.oulu.fi (Oulu NM), http://pgia.ru/CosmicRay/ (Apatity),
 Neutron Monitor Database (NMDB) and IZMIRAN Cosmic Ray database (http://cr0.izmiran.ru/common/links.htm).
NMDB database (www.nmdb.eu), founded under the European Union's FP7 programme (contract no. 213007), is not responsible
 for the data quality.
PIs and teams of all the ballon- and space-borne experiments as well as ground-based neutron monitors whose data
 were used here, are gratefully acknowledged.
This work was partially supported by the ReSoLVE Centre of Excellence (Academy of Finland, project no. 272157).
A.G. acknowledges The Polish National Science Centre, decision number DEC-2012/07/D/ST6/02488.
\end{acknowledgments}


\end{article}


\end{document}